\documentclass[iop]{emulateapj}

\newcommand{\myemail}{\url{bld002@email.uark.edu}}
\newcommand {\apgt} {\ {\raise-.5ex\hbox{$\buildrel>\over\sim$}}\ }
\newcommand {\aplt} {\ {\raise-.5ex\hbox{$\buildrel<\over\sim$}}\ }

\shorttitle{Fundamental Plane of Spiral Galaxies}
\shortauthors{Davis et al.}

\usepackage{color}
\usepackage{soul}
\definecolor{gray}{gray}{0.5}
\usepackage{url}
\usepackage{apjfonts}
\usepackage[latin1]{inputenc}
\usepackage{tikz}
\usetikzlibrary{shapes,arrows}
\usepackage{epsfig}
\usepackage{amsmath}
\usepackage{hyperref}

\submitted{Published in The Astrophysical Journal Letters, 802:L13, 2015, March 24}

\begin{document}

\title{A Fundamental Plane of Spiral Structure in Disk Galaxies}

\author{Benjamin L. Davis\altaffilmark{1}, Daniel Kennefick\altaffilmark{1,2}, Julia Kennefick\altaffilmark{1,2}, Kyle B. Westfall\altaffilmark{3,4}, Douglas W. Shields\altaffilmark{1,2}, Russell Flatman\altaffilmark{5}, Matthew T. Hartley\altaffilmark{2,6}, Joel C. Berrier\altaffilmark{7}, Thomas P. K. Martinsson\altaffilmark{8}, and Rob A. Swaters\altaffilmark{9}}
\altaffiltext{1}{Arkansas Center for Space and Planetary Sciences, University of Arkansas, 346 1/2 North Arkansas Avenue, Fayetteville, AR 72701, USA; \myemail}
\altaffiltext{2}{Department of Physics, University of Arkansas, 226 Physics Building, 835 West Dickson Street, Fayetteville, AR 72701, USA}
\altaffiltext{3}{Kapteyn Astronomical Institute, University of Groningen, P. O. Box 800, 9700 AV Groningen, The Netherlands}
\altaffiltext{4}{Institute of Cosmology and Gravitation, University of Portsmouth, Dennis Sciama Building, Burnaby Road, Portsmouth PO1 3FX, UK}
\altaffiltext{5}{School of Physics, Georgia Institute of Technology, 837 State Street, Atlanta, GA 30332, USA}
\altaffiltext{6}{Physics Department, California Institute of Technology, 1200 East California Boulevard, Pasadena, CA 91125, USA}
\altaffiltext{7}{Department of Physics and Astronomy, Rutgers, The State University of New Jersey, 136 Frelinghuysen Road, Piscataway, NJ 08854-8019, USA}
\altaffiltext{8}{Leiden Observatory, P. O. Box 9513, NL-2300 RA Leiden, The Netherlands}
\altaffiltext{9}{National Optical Astronomy Observatory, 950 North Cherry Ave., Tucson, AZ 85719, USA}

\begin{abstract}

Spiral structure is the most distinctive feature of
disk galaxies and yet debate persists about which
theory of spiral structure is correct. Many versions of the density wave theory demand that the pitch angle
be uniquely determined by the distribution of mass
in the bulge and disk of the galaxy.
We present evidence that the tangent of the
pitch angle of logarithmic spiral arms in
disk galaxies correlates strongly with the
density of neutral atomic hydrogen in the disk and with the central stellar bulge mass of
the galaxy. These three quantities, when plotted against
each other, form a planar relationship that
we argue should be fundamental to our understanding
of spiral structure in disk galaxies. 
We further argue that any successful theory of
spiral structure must be able to explain this
relationship.

\end{abstract}

\keywords{galaxies: evolution --- galaxies: spiral --- galaxies: structure  --- galaxies: fundamental parameters --- galaxies: kinematics and dynamics}

\section{Introduction}

Spiral structure is a commonplace and visually striking feature of many galaxies
and yet there
is still disagreement as to the correct theory that explains its origin after decades of debate. The
first well known theory \citep{Lin:Shu:1964} is that density waves propagating
through the disk of the galaxy are the responsible
agent. The density wave theory for spiral modes, described in detail by \citet{Bertin:Lin:1996}, calls for a long-lived, quasi-steady global spiral pattern. Others call for more transient spiral patterns, whether from swing-amplified noise \citep{Goldreich:Lynden-Bell:1965,Julian:Toomre:1966}, recurrent cycles of groove modes \citep{Sellwood:2000}, or superposed transient instabilities \citep{Sellwood:Carlberg:2014}. Other theories have also been proposed, with
one in particular, the manifold theory \citep{Athanassoula:2009a,Athanassoula:2009b,Athanassoula:2010}, rejecting the density wave concept
altogether in favor of an explanation involving stars in chaotic highly
eccentric orbits.

The density wave theory, as originally articulated by \citet{Lin:Shu:1966}, had a
very specific prediction for the pitch angle of the spiral pattern produced
by the waves.
They calculated the pitch angle to be a ratio of the density of material
in the galaxy's disk to a certain quantity made up 
of the frequencies of orbital motions in the disks, 
\begin{equation}
\tan|P| = \frac
{2\pi m G(\sigma_o+F \sigma_*)}
{R(\kappa^2-(\omega-m\Omega)^2)},
\end{equation}
where $P$ is the logarithmic spiral arm pitch angle, $m$ is the number
of spiral arms in the pattern, $G$ is the gravitational
constant, $\Omega$ is the angular frequency of orbits of particles in the
disk, $R$ is the galactocentric radius, $\kappa$ is the epicyclic frequency of the same particles, and $\omega$
is the frequency associated with the perturbation that excites the density
waves. Note that as long as this perturbation is some form of self-excitation
arising from within the disk itself, it follows that all of the terms in
the denominator should depend on
the mass of the central bulge of the galaxy. The simplest case of a dominant bulge (approximated as a point mass) would imply, for instance,
$\Omega \propto \sqrt{M}$ and $\kappa \propto \sqrt{M}$, with $M$ being the central mass. The numerator depends on the density of gas in the disk, $\sigma_o$, and the stellar disk density, $\sigma_*$, with a factor
$F$, called the reduction factor, which underweights the stellar density (compared
to the gas density), since it is primarily within the gas that the
density wave propagates. In this Letter we will present some evidence that $F << 1$. 

Focusing on the masses and densities involved in this relation, we find
that
\begin{equation}
\tan|P| \propto \frac{\sigma_o+F \sigma_*}{M_o},
\label{DWT}
\end{equation}
where $M_o$ is the mass of the galactic bulge, or else the total mass
interior to the radius in question.

This formula is known to work very well in the case of spiral density
waves in Saturn's rings \citep{Shu:1984}.
Bulge-dominated galaxies are not too distant from the Saturnian
situation of a 
small dense core with negligible mass in the disk, though disk-dominated
galaxies are obviously far more complex. 
Generally speaking, the density wave theory predicts
that the pitch angle of the spiral arms in galaxies does
depend on the radial distribution of matter in the galaxy,
and experimental studies concur \citep{Seigar:2006,Seigar:2014}. Results of
this type broadly agree with \citet{Lin:Shu:1966} that a thin (dense) 
disk and massive (small) central bulge
should result in a tight (loose) spiral. This is not surprising, since we 
would expect a standing
wave pattern (such as in a vibrating string) to depend on the ratio of a 
restoring force or tension (in this case
the central mass, or at least the mass inside a given radius $R$) 
to the density of the medium (in this case the density of
gas in the disk at radius $R$). 
Although the precise nature of the relation between these
three quantities can be expected to vary between galaxies of different
types (bulge-dominated versus disk-dominated, for instance), nevertheless
we show in this Letter that the three quantities, spiral arm pitch angle,
central bulge mass, and gas density in the disk, do strongly correlate to
form a fundamental plane that may play a similar role in tying together
gross features of disk galaxies to that played by the fundamental plane
of elliptical galaxies \citep{Djorgovski:Davis:1987,Dressler:1987}.

We take as our sample disk galaxies from the DiskMass Survey \citep[DMS;][]{DMSI}, which
is ideal for our purposes since it deals with the disk densities of
a sample of face-on galaxies and includes measurements of the central bulge mass.
Using the technique of \citet{Me:2012}, we measure the
pitch angle for these galaxies and find that our sample of
24 galaxies, when plotted in a volume defined by these three
quantities, delineates a plane with very low
scatter. There is only a 0.0047\% chance that this plane could
have been formed by statistical accident.

The plane satisfies a number of requirements, which one would expect of a
useful fundamental plane. The plane is steeply
inclined across the volume formed by the three related quantities. In other
words, it is not merely a relation between two of the three quantities,
with the third essentially irrelevant. The galaxies are distributed quite 
widely and fairly uniformly across the plane. There is no particular 
evidence of a favored curvilinear relation on the face of the plane. 
Finally, and most importantly,
the plane is oriented as one would expect on the basis of the density wave
theory. A large bulge mass and a rarified disk produces the tightest
spirals. A small bulge and a dense disk produce the loosest spirals.
We submit that any successful theory of galactic spiral structure must
be able to explain this result.

One final point is worthy of note. The DMS measured not only the
density of atomic hydrogen in the disk of each galaxy (the quantity
used in our relation), but also the density of molecular hydrogen and
the dynamical disk mass density (the total density in the disk). Our
results suggest that it is the gas density, not the total density in
the disk that matters for spiral density waves. This suggests an apparent decoupling between the stars and gas. The fact that the 
density of atomic hydrogen fits noticeably better than that for molecular
hydrogen may simply be due to the fact
that it is a much more reliable measurement, since molecular hydrogen
is estimated indirectly from observations of other molecules, not
hydrogen itself \citep{Westfall:2011,DMSVII}.

\section{Data and Analysis}

The DMS PPak Sample \citep{DMSVI} consists of 30 nearly face-on galaxies whose disk densities have previously
been closely studied. However, four of these galaxies do not have
central stellar bulge masses available and three provide them only as upper limits, so they are excluded from our sample. In addition to the 23 remaining DMS galaxies, we also include our own Galaxy, the Milky Way, in our sample of galaxies. 


In addition to the method described in \citet{Me:2012}, which utilizes a two-dimensional fast Fourier transform software called \textit{2DFFT}, we also measured pitch angles for all of the sample galaxies using new software called \textit{Spirality}. \textit{Spirality} (Shields et al. 2015, in preparation) is a novel method for measuring spiral arm pitch angle by fitting galaxy images to spiral coordinate systems (templates) of known pitch. For a given pitch angle template, the mean pixel value is found along each of typically 1000 spiral axes. The fitting function, which shows a local maximum at the best-fit pitch angle, is the variance of these means. In other words, we choose the pitch angle that exhibits the greatest contrast between the mean luminosity along the spiral axes. The presumption is that where the pitch angle of the spiral axes is equal to the pitch angle of the galaxy's spiral arms there will be some axes that fall precisely along the true spiral arms (and thus are much brighter in the mean) and some that never coincide with the true spiral arms (and thus are, on average, dim). Where the pitch of the axes is not equal to the pitch of the spiral arms, each axis will cross the true spiral arms a roughly equal number of times, making the mean brightness along each axis roughly equal. Error bars are found by varying the inner radius of the measurement annulus and finding the standard deviation of the best-fit pitch angles.

The two techniques yield measurements that agree within
the error bars in almost all cases. As a final and important test, we
visually inspected each galaxy, comparing them to overlays
of synthetic spirals on transparency paper, in order to confirm
the measured pitch angle. Our overlay transparencies showed
spirals of different sizes and different pitch angles in steps of
$5^{\circ}$ from $5^{\circ}$ to $85^{\circ}$. We were therefore
able to visually confirm the pitch angle to within $5^{\circ}$.
We were satisfied in all cases that the measured pitch angle of
\textit{2DFFT} was reliable and strongly supported by the combination
of \textit{Spirality} and
visual inspection. 
For the sake of consistency, we chose to use
only the results of \textit{2DFFT} in this Letter. The pitch angles, $P$,
given in Table \ref{Sample} are the results of the \textit{2DFFT} routine.

The images used were obtained from the NASA/IPAC Extragalactic Database,\footnote{\url{http://ned.ipac.caltech.edu/}} and/or from the pODI (partial One Degree Imager) camera on the WIYN 3.5 $m$ telescope. The WIYN images were all acquired as 120 $s$ exposures, calibrated using QuickReduce1.0 from the ODI Pipeline, Portal, and Archive,\footnote{\url{http://portal.odi.iu.edu}} and processed using a five-point dither pattern for each galaxy and subsequently stacking the images using \textit{SWarp} \citep{Bertin:2002}. Additionally, KPNO 2.1 $m$ imaging for UGC 463, 1529, 1908, 4036, and 11318 were measured to confirm previous pitch angle measurements. Unless otherwise specified (Milky Way data have been determined in very different ways than other galaxies), all data for stellar galactic bulge 
masses $(M_{\star}^{bulge})$ and maximum neutral atomic hydrogen 
($\rm{H_{I}}$) gas mass surface densities ($\Sigma_{H_{I}}^{max}$) come 
from \citet{Martinsson:2011} and \citet{DMSVII}; see Table \ref{Sample}. For the determination of the stellar bulge masses, the $K$-band light profile was decomposed into a central S\'ersic component (convolved with a seeing disk) and a number of exponential disks \citep{DMSVI}. The bulge masses were determined using the integral of the light from the central S\'ersic component and the mass-to-light ratio ($M/L$) derived from the disk using vertical velocity dispersions \citep{DMSVII}. Gas densities were determined from 21 cm line measurements \citep{Martinsson:2011}.


\begin{deluxetable*}{llccclllc}
\tablecolumns{9}
\tablecaption{Sample\label{Sample}}
\tablehead{
\colhead{Galaxy Name} & \colhead{Type} & \colhead{Band} & \colhead{Image Source} & \colhead{$m$} & \colhead{$\tan|P|$} & \colhead{$\log(M_{\star}^{bulge}/M_{\sun})$} & \colhead{$\Sigma_{H_{I}}^{max}/(M_{\sun}pc^{-2})$} & \colhead{Excluded}  \\
\colhead{(1)} & \colhead{(2)} & \colhead{(3)} & \colhead{(4)} & \colhead{(5)} & \colhead{(6)} & \colhead{(7)} & \colhead{(8)} & \colhead{(9)}
}
\startdata
Milky Way & SBc & 21 cm & 1 & 4 & $0.414\pm0.051$\tablenotemark{a} & $9.95\pm0.03$\tablenotemark{b} & $4.98\pm0.53$\tablenotemark{c,d} &  \\
UGC 448 & SABc & $r$ & 2 & 4 & $0.327\pm0.033$ & $9.76_{-0.51}^{+0.23}$ & $4.58\pm0.46$ &  \\
UGC 463 & SABc & $B$ & 3 & 3 & $0.412\pm0.066$ & $9.35_{-0.41}^{+0.21}$ & $6.18\pm0.66$ &  \\
UGC 1081 & SBc & $r$ & 2 & 2 & $0.452\pm0.064$ & $8.81_{-0.24}^{+0.16}$ & $6.25\pm0.62$ &  \\
UGC 1087 & Sc & $r$ & 2 & 2 & $0.188\pm0.039$ & $8.64_{-0.25}^{+0.16}$ & $4.42\pm0.46$ &  \\
UGC 1529 & Sc & 645.0 nm\tablenotemark{e} & 4 & 3 & $0.490\pm0.096$ & $8.98_{-0.39}^{+0.20}$ & $6.59\pm0.66$ &  \\
UGC 1635 & Sbc & $r$ & 2 & 3 & $0.209\pm0.014$ & $8.74_{-0.29}^{+0.17}$ & $2.60\pm0.32$ &  \\
UGC 1862 & SABcd$^1$ & $r$ & 2 & 2 & $0.444\pm0.074$ & \nodata & $9.14\pm0.91$ & \checkmark  \\
UGC 1908 & SBc$^2$ & 645.0 nm\tablenotemark{e} & 4 & 3 & $0.376\pm0.069$ & $9.68_{-0.54}^{+0.23}$ & $4.62\pm0.46$ &  \\
UGC 3091 & SABd & $i$ & 2 & 2 & $0.555\pm0.092$ & \nodata & $5.59\pm0.56$ & \checkmark  \\
UGC 3140 & Sc & $r$ & 2 & 3 & $0.290\pm0.090$ & $9.65_{-0.24}^{+0.15}$ & $4.87\pm0.54$ &  \\
UGC 3701 & Scd & $r$ & 2 & 2 & $0.276\pm0.090$ & $8.69_{-0.31}^{+0.18}$ & $5.55\pm0.57$ &  \\
UGC 3997 & Im & $g$ & 5 & 2 & $0.185\pm0.048$ & $8.53_{-0.27}^{+0.17}$ & $5.01\pm0.54$ &  \\
UGC 4036 & SABbc & 645.0 nm\tablenotemark{e} & 4 & 2 & $0.268\pm0.021$ & $8.92_{-0.23}^{+0.15}$ & $5.20\pm0.56$ &  \\
UGC 4107 & Sc & $g$ & 5 & 2 & $0.371\pm0.041$ & $8.65_{-0.31}^{+0.18}$ & $5.42\pm0.54$ &  \\
UGC 4256 & SABc & $g$ & 5 & 2 & $0.555\pm0.099$ & $9.29_{-9.29}^{+0.36}$ & $9.75\pm0.98$ & \checkmark  \\
UGC 4368 & Scd & $g$ & 5 & 2 & $0.439\pm0.043$ & $9.21_{-0.41}^{+0.21}$ & $5.95\pm0.66$ &  \\
UGC 4380 & Scd & $g$ & 5 & 3 & $0.430\pm0.095$ & $8.86_{-0.20}^{+0.13}$ & $4.08\pm0.41$ &  \\
UGC 4458 & Sa & $g$ & 5 & 1 & $0.243\pm0.056$ & $10.67_{-0.39}^{+0.20}$ & $3.28\pm0.53$ &  \\
UGC 4555 & SABbc & $g$ & 5 & 2 & $0.213\pm0.017$ & $8.96_{-0.39}^{+0.20}$ & $4.58\pm0.47$ &  \\
UGC 4622 & Scd & $g$ & 5 & 4 & $0.401\pm0.099$ & $9.89_{-0.41}^{+0.21}$ & $3.50\pm0.38$ &  \\
UGC 6903 & SBcd & $g$ & 5 & 2 & $0.283\pm0.041$ & $8.03_{-0.62}^{+0.25}$ & $4.94\pm0.59$ &  \\
UGC 6918 & SABb$^3$ & F606W & 6 & 3 & $0.306\pm0.044$ & $8.04_{-8.04}^{+0.70}$ & $7.04\pm0.72$ & \checkmark  \\
UGC 7244 & SBcd & $g$ & 5 & 2 & $0.627\pm0.105$ & \nodata & $5.53\pm0.60$ & \checkmark  \\
UGC 7917 & SBbc & $g$ & 5 & 3 & $0.278\pm0.025$ & $10.01_{-10.01}^{+00.34}$ & $2.70\pm0.28$ & \checkmark  \\
UGC 8196 & Sb & $g$ & 5 & 5 & $0.144\pm0.009$ & $10.73_{-0.26}^{+0.16}$ & $2.74\pm0.28$ &  \\
UGC 9177 & Scd & $g$ & 5 & 2 & $0.256\pm0.035$ & $9.55_{-0.58}^{+0.24}$ & $3.92\pm0.42$ &  \\
UGC 9837 & SABc & $g$ & 5 & 6 & $0.482\pm0.061$ & $8.35_{-0.29}^{+0.17}$ & $7.95\pm0.80$ &  \\
UGC 9965 & Sc & $g$ & 5 & 3 & $0.237\pm0.037$ & \nodata & $5.63\pm0.58$ & \checkmark  \\
UGC 11318 & SBbc & 645.0 nm\tablenotemark{e} & 4 & 3 & $0.569\pm0.101$ & $9.69_{-0.50}^{+0.23}$ & $6.51\pm0.67$ &  \\
UGC 12391 & SABc & $r$ & 2 & 4 & $0.235\pm0.091$ & $8.98_{-0.28}^{+0.17}$ & $4.90\pm0.49$ &  \\
\enddata
\tablecomments{\emph{Columns:}
(1) Galaxy name.
(2) Hubble morphological type from either the UGC \citep{Nilson:1973} 
or RC3 \citep{RC3} 
catalogs. Notes on morphologies: 1 = peculiar, 2 = starburst, and 3 = AGN.
(3) Filter waveband/wavelength used for pitch angle calculation.
(4) Telescope/literature source of imaging used for pitch angle calculation.
(5) Harmonic mode (number of spiral arms).
(6) Tangent of the pitch angle of the galactic logarithmic spiral arms.
(7) Base 10 logarithm of the stellar bulge mass of the galaxy, in solar masses.
(8) Maximum surface density in the galactic $\rm{H_{I}}$ gas, in solar masses per square pc.
(9) Indication of galaxies that are excluded in fittings due to missing measurements or measurements that are merely upper limits.
\emph{Image Sources:}
(1) \citet{Levine:2006};
(2) WIYN 3.5 $m$ pODI;
(3) JKT 1.0 $m$;
(4) Palomar 48 inch Schmidt;
(5) SDSS;
(6) HST.
}
\tablenotetext{a}{\citet{Levine:2006}.}
\tablenotetext{b}{\citet{McMillan:2011}.}
\tablenotetext{c}{No error estimates were provided by its reference so we have assigned the mean error of the included sample, $\pm0.53$ $M_{\sun}pc^{-2}$.}
\tablenotetext{d}{Calculated using Equation (2) from \citet{Ferriere:2001}.}
\tablenotetext{e}{IIIaJ emulsion.}
\end{deluxetable*}

\vspace{9mm}

\section{Results}

\begin{figure}
\includegraphics[width=8.6cm]{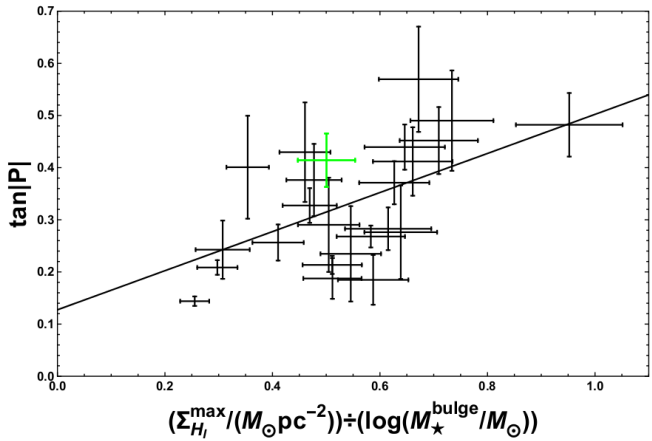}
\caption{Two-dimensional plot of the linear fit defined by the multivariate normally distributed sampling fit of Equation (\ref{propto}) depicted by the solid black line, along with the plotted points of the 24 galaxy member data set. The Milky Way is depicted distinctly in \textcolor{green}{green}. The axes $[x, y]$ depict [$\tan|P|$, $(\Sigma_{H_{I}}^{max}/(M_{\sun}pc^{-2}))/\log(M_{\star}^{bulge}/M_{\sun})$], respectively.
\label{2DPlot}}
\end{figure}

We find a best linear fit for Equation (\ref{DWT}) from the included data sample of 24 galaxies of 
\begin{equation}
\tan|P| = (0.375\pm0.092)\frac{\Sigma_{H_{I}}^{max}/(M_{\sun}pc^{-2})}{\log\left(M_{\star}^{bulge}/M_{\sun}\right)} + (0.127\pm0.049).
\label{propto}
\end{equation}
The root mean squared error ($E_{rms}$) is equal to 0.0909 (a residual scatter of $31.2\%$ per galaxy on average), with $R^2 = 0.344$, and a $p$-value equal to $2.59$ $\times$ $10^{-3}$ for Equation (\ref{propto}). A plot of this linear fit, along with the included data sample, is given in Figure \ref{2DPlot}.


\begin{figure*}
\includegraphics[width=5.97cm]{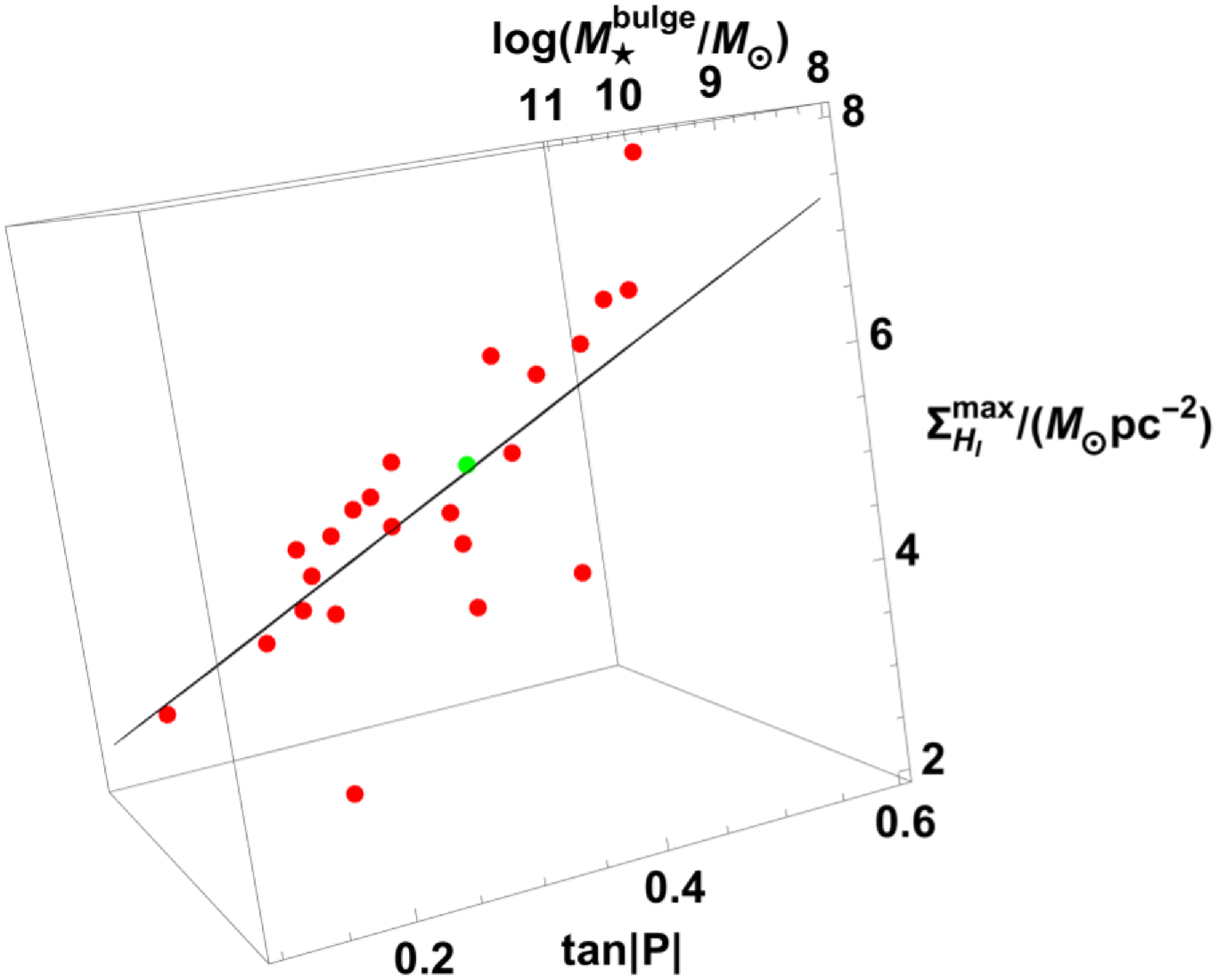}
\includegraphics[width=5.97cm]{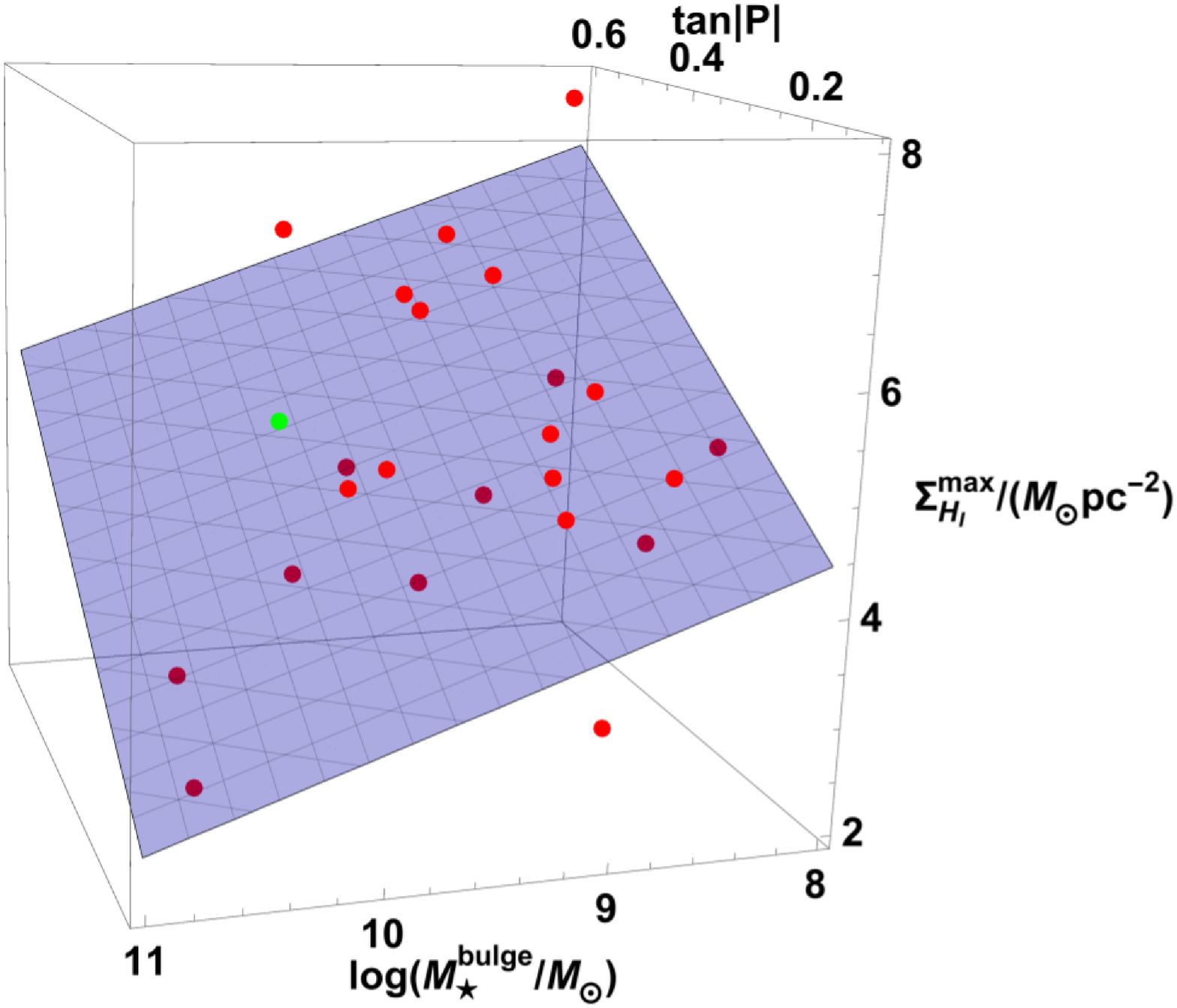}
\includegraphics[width=5.97cm]{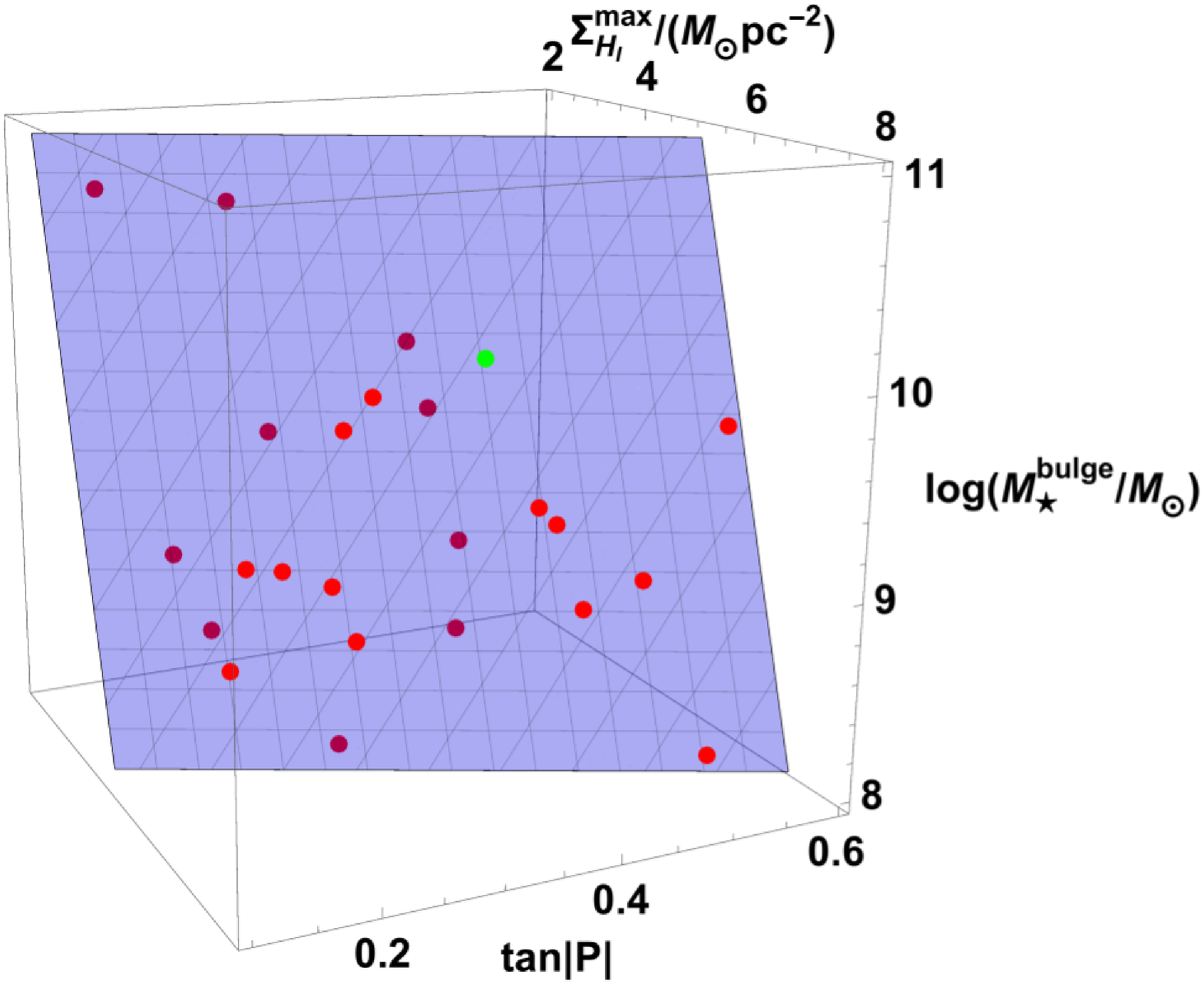}
\caption{Three-dimensional plot of the plane defined by the fit of Equation (\ref{Plane}) with the multivariate normally distributed sampling depicted by a translucent \textcolor{blue}{blue} meshed surface, along with the plotted points of the 24 galaxy member included data set (depicted by \textcolor{red}{red} spheres with the Milky Way in \textcolor{green}{green}). Note that the points will appear slightly darker when they are projected behind and partially obscured by the plane. The axes $[x, y, z]$ depict [$\tan|P|$, $\log(M_{\star}^{bulge}/M_{\sun})$, $\Sigma_{H_{I}}^{max}$], respectively. Left: the view has been oriented parallel to the plane. Middle: the view has been oriented at an orientation sufficient to view the face of the plane. Right: the view has been projected along an orthogonal vector above the plane.\label{3DPlot}}
\end{figure*}

The formula describing the fundamental plane for spiral galaxies from the sample is as follows:
\begin{eqnarray}
\frac{\Sigma_{H_{I}}^{max}}{M_{\sun}pc^{-2}} = (5.70\pm1.40)\tan|P| - \nonumber  \\
(0.677\pm0.199)\log\left(M_{\star}^{bulge}/M_{\sun}\right) + (9.29\pm1.96).
\label{Plane}
\end{eqnarray}
$E_{rms} = 0.770$ $M_{\sun}pc^{-2}$ (a residual scatter of $16.7\%$ per galaxy on average\footnote{Compare this to the fundamental plane for elliptical galaxies, which has a residual scatter of $\sim20\%$ per galaxy on average \citep{Kormendy:Djorgovski:1989}.}), with $R^2 = 0.613$, and a $p$-value $= 4.71$ $\times$ $10^{-5}$ for Equation (\ref{Plane}). It is interesting to note that the addition of the extra dimension cuts the residual scatter approximately in half. A three-dimensional plot of this plane, along with the included data sample, is given in Figure \ref{3DPlot}\footnote{A 3D animated gif of this figure can be accessed at \url{http://dafix.uark.edu/~ben/movie.gif}.}.


The errors presented in Equations (\ref{propto}) and (\ref{Plane}) have been determined by sampling individual data points from multivariate normal distributions using the following algorithm.
\begin{enumerate}
	\item For each measurement, draw a new measurement based on multivariate normal distributions with the mean and variance of each variable for all 24 galaxies. 
	\item Fit linear (or planar) best-fit coefficients to the 24 points.
	\item Repeat steps 1 and 2 $10^6$ times, saving the fitted coefficients after each step.
	\item Use the distribution of the coefficients based on these $10^6$ fits to define the best-fitting (median) coefficients and their error (the $1\sigma$ confidence interval of the distribution).
\end{enumerate}

The orientation of the fundamental plane illustrated in Figure \ref{3DPlot} is exactly as one would expect on the basis of the spiral density wave theory. According to Equation (\ref{DWT}), the pitch angle is minimized (tightest winding) when the $\rm{H_{I}}$ mass surface density is low and the central mass is high. 
Alternatively, the pitch angle is maximized (loosest winding) when the $\rm{H_{I}}$ mass surface density is high and the central mass is low. This behavior is illustrated in both figures. Note in the middle panel of Figure \ref{3DPlot} how the plane slopes from the lower left front (low pitch angle, low $\rm{H_{I}}$ mass surface density, and high central mass) corner of the cube to the upper right back (high pitch angle, high $\rm{H_{I}}$ mass surface density, and low central mass) corner of the cube. 
Furthermore, this indicates that the shape of the plane is strongly correlated to all three variables (the individual variable $p$-values of the intercept,  $\tan|P|$, and $\log(M_{\star}^{bulge}/M_{\sun})$ are $9.97$ $\times$ $10^{-5}$, $8.03$ $\times$ $10^{-5}$, and $9.41$ $\times$ $10^{-4}$, respectively for Equation (\ref{Plane})).

\section{Discussion}

If one favors the standing wave picture of spiral structure, our result
is not unexpected.
In analogy with standing waves on a string, we would expect the
wavelength of the resulting pattern to be strongly determined by the
tension or restoring force (in this case the central gravitating mass)
and the density of the medium (in this case the gas in the disk).
It is worth noting that our case is probably analogous to a string with
non-uniform density, since the gas density generally falls off with increasing radius in a
galactic disk\footnote{This is true for the total gas density; however, the
atomic gas density generally has a peak value at some radius and decreases
toward the center due to conversion of the atomic to molecular gas.}. Additionally, it seems reasonable that
the gravitational restoring force increases
with increasing radius, since there will be more mass inside the given
radius. Both effects would tend to cause the pitch angle to tighten with
increasing radius, and this effect is often seen in spiral galaxies \citep[see, e.g.,][]{Me:2012}.

It has been often proposed in the past that different mechanisms may
explain spiral structure in different galaxies; for instance, the
mechanism that produces grand design spirals may differ from the one
that produces flocculent spirals \citep[e.g.,][]{D'Onghia:2013}. Furthermore, it has been demonstrated that galaxies that appear to have grand design structure in infrared light can appear flocculent in blue images, suggesting that stellar and gaseous disks are decoupled \citep{Grosbol:Patsis:1998}. It is striking that the sample used
in this study contains quite a few flocculent or multi-armed patterns,
which are not grand design. The existence of a very low
scatter planar correlation for all of these galaxies is thus very
significant and implies that different galactic morphologies all adhere to the same imposed mechanism of density wave theory. It is true that four of the galaxies have a noticeably
greater scatter than the others, and further study with larger
samples might yet support the existence of two kinds of spiral
structure. It is worth noting that two of these four galaxies represent
the extremes of gas density for the sample, one having clearly the
highest gas density in its disk, another clearly the lowest.


In recent years, there has been some discussion that spiral arms may be quite transient, persisting for only one or two revolutions of the disk galaxy \citep{Toomre:Kalnais:1991}. In recent years, there have been attempts to show theoretically
that more long-lasting spiral patterns are possible \citep{Sellwood:Carlberg:1984,D'Onghia:2013}. This Letter
suggests that even if spiral patterns are transient,
some resonant mechanism compels the pattern, when it reforms, to resume something close to its previous pitch angle.

The relation discovered here might be useful as a tool in the study
of disk galaxies. One of the three quantities, disk density, is
relatively difficult to measure. The relation found here could be
used to measure it indirectly from the other two quantities
(pitch angle and central bulge mass), which would be easier to
measure. In addition, the existence of the three-way correlation
may enable more careful studies of the important relation between
pitch angle and central mass, which is itself a very useful marker
for quantities such as the central black hole mass \citep{Seigar:2008,Berrier:2013}.

It has long been known that pitch angle does depend on the distribution
of mass \citep[e.g.,][]{Seigar:2006,Seigar:2014}
and on the size of the central bulge
(for instance, the observed correlation between pitch angle and sigma
reported in \citet{Seigar:2008} and \citet{Berrier:2013}, as well as the qualitative relation of
pitch angle to bulge size featured in the Hubble classification).
In addition, it has been reported that pitch angle varies with the
total mass of gas in galactic disks \citep{Roberts:1975}.
This Letter demonstrates the fundamental way in which we can understand
how the spiral structure depends on the distribution of mass in
disk galaxies. Furthermore, it illustrates how the qualitative Hubble morphological types can exhibit varying pitch angles for galaxies that have similarly sized bulges and are thus categorized as the same type. These galaxies likely have different gas densities in their disks. Although the density wave theory provided the
inspiration for this study, other theories may also be able to
explain this result. Certainly, any successful theory of spiral
structure must be able to do so.

\acknowledgments

The authors thank Marc Seigar and Matthew Bershady, who provided valuable comments on drafts of the Letter. We thank the National Optical Astronomy Observatory for observing time on the WIYN 3.5 $m$ telescope at Kitt Peak National Observatory. We acknowledge the use of data obtained with PPak and SparsePak IFUs on the Calar Alto and WIYN 3.5 $m$ telescopes, made available by the DiskMass Survey team. This  research has  made  use of  the 
NASA/IPAC Extragalactic  Database, which  is operated by  
the Jet Propulsion  Laboratory,  California  Institute  of  Technology,  
under contract with the National Aeronautics and Space Administration. 
Plots were generated using \textit{Mathematica}. Numerical computations and statistical analyses were performed with \textit{Matlab}. We also thank 
the National Science Foundation (NSF) for REU Site grant \#0851150, which 
contributed to a significant part of the data collection for this Letter. This research has made use of NASA's Astrophysics 
Data System. K. B. W. acknowledges support from NSF (USA) grant OISE-0754437 and NWO (NL) grant 614.000.807.

\end{document}